\documentclass[%
 reprint,amsmath,amssymb,pra,aps]{revtex4-1}
                                                                                                                                                                                                                                                                                                                                                                                                                                                                                                                                                                                                                                                                                                                                                                                   \usepackage{graphicx}
\usepackage{dcolumn}
\usepackage{bm}
\usepackage{braket}
\usepackage{epstopdf}
\usepackage[sort&compress]{natbib}
\begin{document}
\title{Motion-induced enhancement of Rabi coupling between atomic ensembles in cavity optomechanics}
\author{Anil Kumar Chauhan}
\author{Asoka Biswas}%
\email{abiswas@iitrpr.ac.in}
\affiliation{%
Department of Physics, Indian Institute of Technology Ropar, Rupnagar, Punjab 140001, India
}%
\date{\today}%

\begin{abstract}
We propose a scheme of enhancement of Rabi coupling between two identical atomic ensembles trapped inside an optical cavity in a membrane-in-the-middle set up. The cavity modes dispersively interact with the ensembles and the effective interaction between the ensembles is governed by the tunnelling rate of the cavity modes through the oscillating membrane. We have shown that this interaction can be made large enough such that the Rabi oscillation occurs in a time-scale, much smaller than the relevant decay time-scales of the cavity modes and of the membrane. We present the detailed analytical and numerical results and assess the feasibility of the scheme using currently available technology. 
\end{abstract}
\maketitle
\section{Introduction}
Light-matter interaction is in the heart of quantum communication  and  information processing \cite{Gisin2007}. Flying atoms \cite{haroche}, trapped atoms \cite{Kimble2005,englert} and ions \cite{Wineland2003} interacting with cavity modes have been demonstrated as suitable platforms for quantum communication. In such systems, light field carries the information from one node to the other, in which the trapped atoms or ions work as nodes, and store and manipulate the information. However, maintaining the coherence of the information carrier for a long time and the other technical difficulties to strongly couple a single atom with a high-finesse cavity pose severe bottlenecks to realizing quantum communication using such systems \cite{Kimble2008}. The time-scale of such realization can be improved by considering a regime of strong coupling between the atoms and the cavity mode \cite{englert}. This could be achieved by increasing the number of photons inside the cavity, which however leads to a large decay rate and therefore to faster decay of coherence. A suitable better way of obtaining strong coupling is to use an atomic ensemble trapped inside the cavity. For $N (\gg 1)$ number of atoms in such an ensemble, the coupling increases as $\sqrt{N}$, in low atomic excitation limit. Further, in such a limit, the ensemble can also serve as a storage of quantum information, as the ground states of the atoms are immune to spontaneous emission.  A nice review on the interface between the atomic ensembles and the light field can be found in \cite{Klemens2010}. Several models of ultracold atomic gases in optical lattices and their application in quantum information processing have been discussed in \cite{sen}. Using the interaction between ultracold atomic ensemble and the quantized light field, one could further attain a new regime of light-mater interaction \cite{ritsch}. 

Recent advances in cavity optomechanics show a possibility of coupling a mesoscopic mechanical oscillator with the cavity mode \cite{Aspelmeyer2014}. Interesting feature of such systems is its hybrid nature, in which one combines `advantages of different physical systems in one architecture'.  Such systems may be useful in coupling the collective spin of an atomic ensemble to the oscillator, mediated by the cavity field (\cite{Klemens2010} and references therein). In this way, an atomic ensemble instead of a single atom may be treated as a suitable platform of quantum information processing, all using mesoscopic systems only. Particularly, ultracold atoms have been quite promising in this regard. Interface between such atoms with the mechanical oscillator and the strategies to enhance their coupling using optical lattice or cavities have been reviewed in \cite{hunger}. One could also achieve an effective coupling, akin to spin-orbit coupling, between the internal degrees of freedom and the center-of-mass motion of these atoms, using cavity modes \cite{dong}. Interestingly, the motion of ultracold quantum gases can be used to replace the role of a mechanical oscillator towards mimicking an equivalent cavity optomechanical system \cite{murch}. 

In this paper, we show that the dynamics of two identical atomic ensembles can be manipulated using a mesoscopic oscillator, rather than using a light field, as is usually done in the relevant experiments \cite{Klemens2010}. We choose a membrane-in-the-middle set up, in which a membrane suspended at a suitable position inside a Fabry-Perot cavity interacts with the cavity mode \cite{bhatta-1} and two identical atomic ensembles, each containing $N$ atoms and each interacting with two different modes of the cavity are trapped in the either side of the membrane. The membrane divides the cavity into two halves, corresponding to two orthogonal modes $a_L$ and $a_R$ of the cavity, which can couple to each other via tunnelling through the membrane \cite{law1,marquardt}.  In such a set up, the strength of coupling between a single cavity mode and the membrane usually becomes a quadratic function of the displacement of the membrane from its equilibrium position \cite{Thompson2008}. However, suitable resonance condition \cite{law1,marquardt,komar} can give rise to a coupling that is linear in displacement and that  arises even when the membrane is placed exactly in the middle of the cavity. 
Since, the cavity decay rate is usually much larger than that of the mechanical system, it is preferable to work in the adiabatic limit, to combat the effect of cavity leakage. 
In this paper, we further use the cavity-laser detuning much larger than the atom-laser detuning, to facilitate the adiabatic elimination of the cavity modes. We present detailed analytic calculation to show that the enhancement of Rabi coupling between the ensembles is indeed possible, governed by the rate of tunnelling through the membrane. 

The paper is organised as follows. We present our model in Sec. II. In Sec.III, we obtain an effective Hamiltonian and discuss how large Rabi coupling can be achieved. We conclude the paper in Sec. IV.   
\section{Model}
We consider an optomechanical cavity setup in which a membrane is suspended inside the cavity either at a node or at an antinode of the cavity mode frequency, as is usually done in a membrane-in-the-middle configuration. The cavity modes on the either side of the membrane are thereby coupled to the membrane through radiation pressure.
\begin{figure}[h]
\includegraphics[width=5cm]{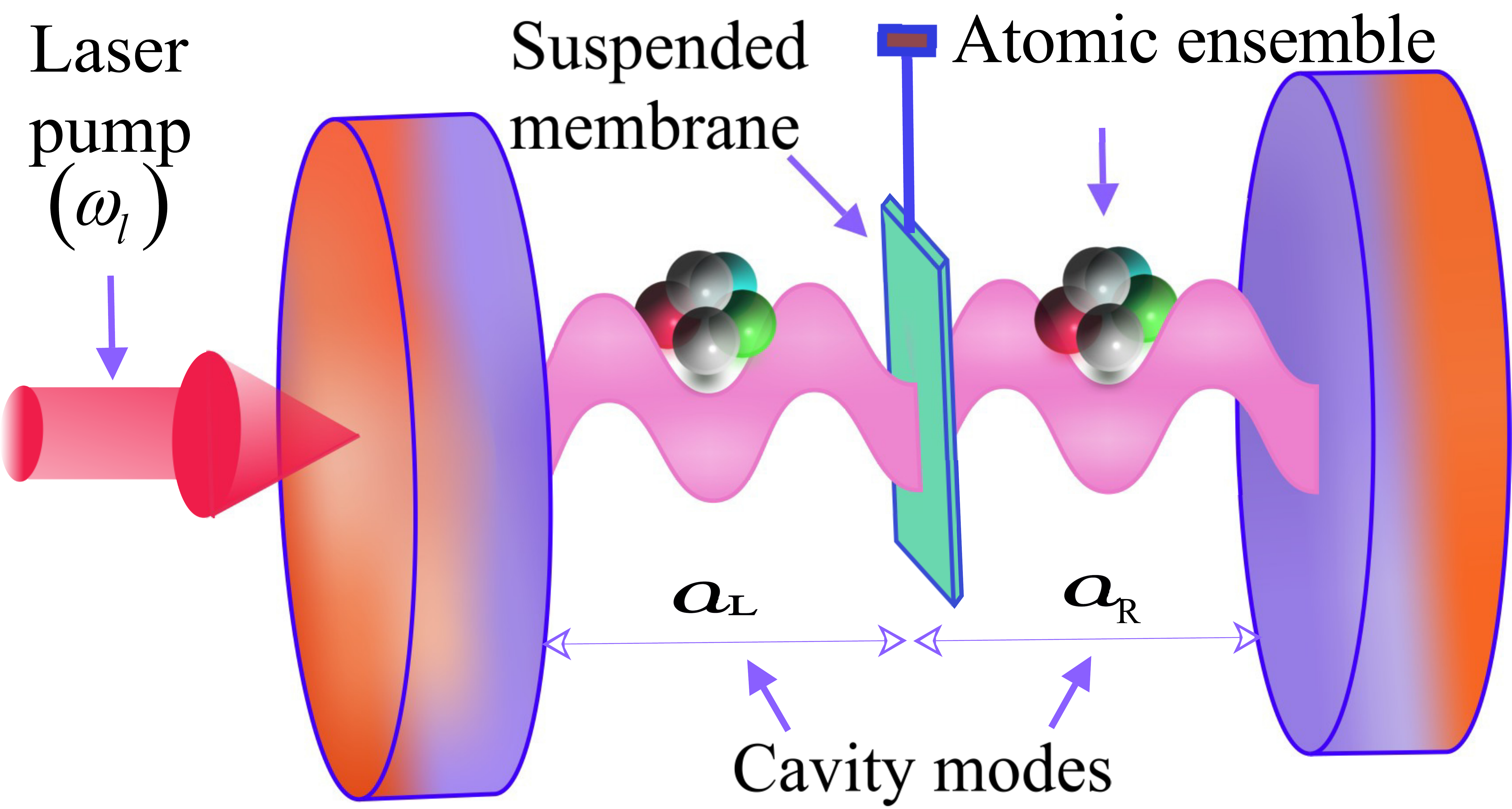}
\centering
\caption[Fig(a)]{Schematic diagram of two trapped atomic ensemble in hybrid optomechanical cavity.}
\end{figure}
We assume that a pump laser is driving the mode on the left half of the cavity. Therefore, the Hamiltonian of the system can be written as follows \cite{law1}:
\begin{eqnarray}
H_{\rm ac}&=&H_0+H_I+H_{\rm p}\;,\nonumber\\ 
H_{0} & =& \sum_{\alpha\in {\rm L,R}} \left(\omega_{c,\alpha} {a_{\alpha}^\dag} a_{\alpha}\right)+   \omega_{m} {b^\dag}b \nonumber\\
H_I&=&-J\left( a_{L}^\dag a_{R}+ {\rm h.c.}\right) -g_{1} \left( a_{L}^\dag a_{L}-a_{R}^\dag a_{R}\right)\left( b+b^\dag\right)\;,\nonumber\\
H_{\rm p} & =&  \epsilon\left( a_{L}^\dag e^{-i\omega_{l}t}+a_{L} e^{i\omega_{l}t}\right)\;,
 \end{eqnarray}
where $H_{0}$ is the bare energy part of the Hamiltonian, $a_L$ and $a_R$ are the annihilation operators for the cavity modes on the left and the right side of the membrane with corresponding frequencies $\omega_{c,L}$ and $ \omega_{c,R}$, respectively, and $b$ is the annihilation operator of the mechanical mode with frequency $\omega_{m}$. In the interaction part $H_{I}$ of the Hamiltonian, $J$ represents the tunnelling rate between the two cavity modes through the membrane and $g_{1}$ is the optomechanical coupling. 
$H_{\rm p}$ represents the pumping of the left cavity mode by a laser field with frequency $\omega_l$ and a coupling strength $\epsilon= \sqrt{2\gamma_c \rm{P}/\omega_l}$, where $P$ is  the power of the driving field and $\gamma_c$ is the cavity decay rate. 

We next consider two identical atomic ensembles, each containing $N$ two-level atoms trapped in the left and the right side of the cavity, respectively (see Fig. 1). These ensembles interact with the respective cavity mode $a_{\rm L,R}$, as described by the following Hamiltonian:
\begin{equation}
H_{\rm at}=\sum_{\alpha\in {\rm L,R}}  \omega_\alpha S^{(\alpha)}_z+\bar{g}\sum_{\alpha\in {\rm L,R}}\left( S_{-}^{(\alpha)}a_{\alpha}^\dag+{\rm h.c.}\right)\;,
\end{equation}
 where $\omega_\alpha$ is the transition frequency of each atom in the ensemble on the $\alpha$-side and the operators $S^{(\alpha)}$ represent the collective spin operators of the $N$ spins on the $\alpha$-side, i.e., $S_z=\sum_{k=1}^N\sigma_z^{(k)}$ for either side of the membrane. Further, $\bar{g}=\sum_{k=1}^{N}g^{k}/N$ is the average atom-cavity coupling for the ensemble comprising $N$ atoms, on the either side of the mechanical oscillator, where $g^{k}$ accounts for the coupling of the $k$th atom with the cavity mode. 
 
For a large number of atoms in the ground state or in low atomic excitation limit, the collective spin operator $S_{\pm}$ and $S_{z}$ of the atomic ensemble can be transformed to an equivalent bosonic operator $c$ and $c^\dag$ through the Holstein-Primakoff representation \cite{Holstein1940,Genes2008} 
\begin{eqnarray}
S_{-}&=&c\sqrt{N-c^\dag c}\simeq\sqrt{N} c,\\
S_{+}&=&c^\dag \sqrt{N-c^\dag c}\simeq \sqrt{N}c^\dag,\\
S_{z}&=& c^\dag c - \frac{N}{2},
\end{eqnarray}
where $N$ is the number of atoms in the ensemble, $ c=S_{-}/\sqrt{|{\left<S_{z}\right>}|}$ and $ c^{\dag}=S_{+}/\sqrt{|{\left<S_{z}\right>}|}$ satisfy the commutation relation $[c,c^{\dag}]=1$. 
Therefore, the transformed Hamiltonian is 
\begin{equation}
H'= H_{\rm ac}+H'_{\rm at}\;,
\end{equation}
where
\begin{eqnarray}
H'_{\rm at} & = &  \sum_{\alpha\in {\rm L,R}}\omega_{\alpha}\left( c^\dag_{\alpha} c_{\alpha} -\frac{N}{2}\right)+\sqrt{N}\bar{g}\left( c_{L}a_{L}^\dag+c_{R}a_{R}^\dag +{\rm h.c.}\right).\nonumber\\
\end{eqnarray}
 In the rotating frame of laser with frequency $\omega_{l}$, the Hamiltonian takes the following form:
 \begin{eqnarray}\label{eq:H}
 H & = & \sum_{\alpha\in{\rm L,R}}\left( \delta_\alpha {a_\alpha^\dag} a_\alpha\right) + \omega_{m} {b^\dag}b +\sum_{\alpha\in {\rm L,R}} \Delta_{\alpha}\left( c^\dag_\alpha c_\alpha -\frac{N}{2}\right)\nonumber \\
& &  + H_{I} + \sqrt{N}\bar{g}\left( c_{L}a_{L}^\dag+c_{R}a_{R}^\dag +{\rm h.c.}\right)+
 \epsilon\left( a_{L}^\dag +a_{L} \right),\nonumber\\
 \end{eqnarray}
where $\delta_{L} =\omega_{c,L}- \omega_{l}$ and $\delta_{R}= \omega_{c,R}-\omega_{l}$ are the detunings of the left and the right cavity modes, respectively, with the pump field, while, $\Delta_\alpha=\omega_{\alpha} - \omega_{l}$ is that of the atomic ensemble on the $\alpha$-side of the membrane with the pump field. 

The above Hamiltonian is nonlinear, as the cavity modes couple to the modes of the membrane through resonant-mode interaction \cite{law1}, governed by the interaction $H_I$. In order to study the fluctuation dynamics of the cavity mode and the oscillators, we use the standard linearization procedure \cite{Aspelmeyer2014}, in which one expands all the bosonic operators as a sum of the average values and the zero-mean fluctuation, as follows: $a\rightarrow \alpha + \delta a$, $b\rightarrow \beta + \delta b$, $c_{L}\rightarrow \xi_{L}+ \delta_{L}$,  $c_{R}\rightarrow \xi_{R}+ \delta c_{R} $. Here $\alpha$, $\beta$, $\xi_{L}$, and $\xi_{R}$ are in general complex and denote the steady state values of the respective annihilation operators of the cavity, membrane, and the collective atomic modes on the left and the right side of the membrane. Applying this transformation to Eq. (\ref{eq:H}), we obtain the following linearized form of the Hamiltonian, as given by 
\begin{eqnarray}{\label{linear}}
 H & = & \sum_{\alpha\in {\rm L,R}} \left[ \delta'_{\alpha} \left( {\delta a_{\alpha}^\dag} \delta a_{\alpha}\right)+ \Delta_{\alpha}\left( \delta c^\dag_{\alpha} \delta c_{\alpha}\right)\right] \nonumber\\
& & +\sqrt{N}\bar{g}\left( \delta c_{L}\delta a_{L}^\dag+ \delta c_{R}\delta a_{R}^\dag +{\rm h.c.}\right)  \nonumber\\
& & -J\left( \delta a_{L}^\dag \delta a_{R}+ {\rm h.c.}\right) + \omega_{m}\delta b^{\dag} \delta b\nonumber\\
& &  -g_{1}\left[ \alpha_{L}\left(\delta a_{L}+\delta a_{L}^{\dag}\right)-\alpha_{R}\left(\delta a_{R}+\delta a_{R}^{\dag}\right)\right]\nonumber\\
 & &\left({\delta b}  +{\delta b^{\dag}}\right) - g_{1}\left( \delta a_{L}^\dag \delta a_{L}-\delta a_{R}^\dag \delta a_{R}\right) \left[{\delta b} +{\delta b^{\dag}} \right],\nonumber\\
\end{eqnarray}  
where $\delta'_L=\delta_L-2g_1\beta$ and $\delta'_R=\delta_R+2g_1\beta$ represent the modified detuning of the cavity mode on the left and the right side of the membrane, respectively, and we have chosen $\alpha_{L}$,  $\alpha_{R}$, and $\beta$ to be real.

The Langevin equations for the fluctuations can then be written, using Eq.(\ref{linear}) and the input-output formalism, described in \cite{Wall2008},  as follows: 
\begin{eqnarray}
\dot{\delta a_{L}} &= & -\left(\frac{\gamma_{L}}{2}+i\delta'_{L}\right)\delta a_{L} - i\bar{g}\sqrt{N} \delta c_{L} + i J \delta a_{R}\nonumber\\
\label{beforead1}& & +ig_1\alpha_{L}\left(\delta b+ \delta b^{\dag}\right)+ \sqrt{\gamma_{L}} a_{L}^{in},\\
 \dot{\delta a_{R}}  & = & -\left(\frac{\gamma_{R}}{2}+i\delta'_{R}\right)\delta a_{R} - i\bar{g}\sqrt{N} \delta c_{R} + i J \delta a_{L}\nonumber\\
& &- ig_1\alpha_{R}\left(\delta b+ \delta b^{\dag}\right) + \sqrt{\gamma_{R}} a_{R}^{in},\\
\dot{\delta b}& = & -\left[\frac{\gamma_{m}}{2}+i\omega_m\right]\delta b \nonumber\\
& & + ig_{1}\left[\alpha_{L}\left(\delta a_{L}+ \delta a_{L}^{\dag}\right)-\alpha_{R}\left(\delta a_{R}+\delta{a_{R}^{\dag}}\right)\right]\nonumber\\
& & +\sqrt{\gamma_{m}} b^{in},\\
 \dot{\delta{c_{L}}} &  = & -\left(\frac{\gamma_{1}}{2}+ i\Delta_{L}\right)\delta c_{L}-i \bar{g}\sqrt{N}\delta a_{L}+\sqrt{\gamma_{1}}c_{L}^{in},\\
\label{beforead2} \dot{\delta{c_{R}}} & = &-\left(\frac{\gamma_{2}}{2}+ i\Delta_{R}\right)\delta c_{R}-i \bar{g}\sqrt{N}\delta a_{R}+\sqrt{\gamma_{2}}c_{R}^{in},
\end{eqnarray}
where $\gamma_{L}$ and $\gamma_{R}$ are the decay rates of the two modes of the cavity, $\gamma_{m}$ is the rate of mechanical dissipation, and $\gamma_{1}$ and $\gamma_{2}$ are the decay rates of the atomic ensembles. The corresponding noise operators $a_{L}^{in}$, $a_{R}^{in}$, $b^{in}$ , $c_{L}^{in}$, and $c_{R}^{in}$ satisfy the following correlations \cite{Wall2008}
\begin{eqnarray}
 \left\langle a_{L}^{in}(t)a_{L}^{\dag in}(t')\right\rangle &= &  \left\langle a_{R}^{in}(t)a_{R}^{\dag in}(t')\right\rangle=\delta(t-t'),\nonumber\\ 
 \left\langle a_{L}^{\dag in}(t)a_{L}^{in}(t')\right\rangle &= &  \left\langle a_{R}^{\dag in}(t)a_{R}^{in}(t')\right\rangle= 0 ,\nonumber\\ 
 \left\langle c_{L}^{in}(t) c_{L}^{in \dag}(t')\right\rangle & = & \left\langle c_{R}^{in}(t) c_{R}^{in \dag}(t')\right\rangle =\delta(t-t'),\nonumber\\
 \left\langle c_{L}^{\dag in}(t) c_{L}^{in}(t')\right\rangle & = & \left\langle c_{R}^{\dag in}(t) c_{R}^{in}(t')\right\rangle =0 ,\nonumber\\
 \left\langle b^{in}(t) b^{in \dag}(t')\right\rangle & = & (\bar{n}_{th}+1)\delta(t-t'),\nonumber\\ 
 \left\langle b^{\dag in}(t) b^{in}(t')\right\rangle & = & \bar{n}_{th}\delta(t-t'),\nonumber 
\end{eqnarray}
where $\bar{n}_{th}=\left\{ \exp\left[\hbar\omega_{m}/(k_{B}T)\right]-1\right\}^{-1}$ is the mean thermal excitation number of the bath mode interacting with the mechanical oscillator with frequency $\omega_{m}$ at an equilibrium temperature $T$ and $k_{B}$ is the Boltzmann constant.

We next choose the decay rates of both the cavity modes to be the same, i.e., $\gamma_{L}=\gamma_{R}=\gamma_c$, such that the Langevin equations for the cavity mode fluctuations become
\begin{eqnarray}
\dot{\delta a_{L}}(t) & = & -\left(\frac{\gamma_c}{2}+i\delta'_{L}\right)\delta a_{L} - i\bar{g}\sqrt{N} \delta c_{L} + i J \delta a_{R}\nonumber\\
& &+ig_1\alpha_{L}\left(\delta b+ \delta b^{\dag}\right) + \sqrt{\gamma_c} a_{L}^{in},\\
\dot{\delta a_{R}}(t) & =& -\left(\frac{\gamma_c}{2}+i\delta'_{R}\right)\delta a_{R} - i\bar{g}\sqrt{N} \delta c_{R} + i J \delta a_{L}\nonumber\\
& &- ig_1\alpha_{L}\left(\delta b+ \delta b^{\dag}\right) + \sqrt{\gamma_c} a_{R}^{in}.
\end{eqnarray}

\subsection{Adiabatic Approximation}
In the limit of large cavity decay rate $\gamma_c\gg \gamma_{\rm at}, \gamma_m$ and large effective cavity-laser detuning $|\delta'_{L,R}|\gg\Delta_{L,R}$ , the dynamics of fluctuations in the cavity modes can be neglected and these modes  can be eliminated adiabatically \cite{lugiato,adia,chen}. In this limit, we therefore substitute $\dot{\delta a_{L}}\approx 0$ and $\dot{\delta a_{R}}\approx 0$, leading to
\begin{eqnarray}\label{eq:al}
\delta a_{L}& =& \frac{1}{z}\left[ (-Jg_1\alpha_{R}+ g_1y\alpha_L)\left(\delta b+\delta b^{\dag}\right)- \bar{g} \sqrt{N} y\delta c_{L}\right.\nonumber\\
& &\left. - \bar{g} \sqrt{N} J\delta c_{R} -i \left( y a_{L}^{in}\right)\sqrt{\gamma_c} -iJ\sqrt{\gamma_c}a_{R}^{in}\right],\nonumber\\
\delta a_{R}& =& \frac{1}{z}\left[ (Jg_1\alpha_{L}- g_1 x\alpha_R)\left(\delta b+\delta b^{\dag}\right)- \bar{g} \sqrt{N} x\delta c_{R}\right.\nonumber\\
& &\left. - \bar{g} \sqrt{N} J\delta c_{L} - i \left( x a_{R}^{in}\right)\sqrt{\gamma_c} -iJ\sqrt{\gamma_c}a_{L}^{in}\right],
\end{eqnarray}
where $x= i\frac{\gamma_c}{2} -\delta'_{L}$, $y= i \frac{\gamma_c}{2}-\delta'_{R}$, and $ z=xy-J^{2}$.
Using the above equations, we obtain the modified Langevin equations as follows:
\begin{eqnarray}
\dot{\delta{b}} &=&  \left(-i\tilde{\omega}_{m}-\frac{\gamma_{m}}{2}\right)\delta b-i\Lambda\delta b^{\dag}+\sqrt{\gamma_{m}}b^{in}\nonumber\\
&&- i \left( G_{\rm{eff}} \delta c_{L}+ {\rm h.c.}\right)+ i \left(\bar{G}_{\rm{eff}} \delta c_{R}+ {\rm h.c.}\right) \nonumber\\
&&+ i g_1 \alpha \sqrt{\gamma_c}\left[\left(i\frac{J}{z}a_{L}^{in} +i\frac{x}{z}a_{R}^{in}\right)- {\rm h.c.}\right]\nonumber\\
\label{b}&&- i g_1 \alpha \sqrt{\gamma_c}\left[\left(i\frac{J}{z}a_{R}^{in}+i\frac{y}{z}a_{L}^{in}\right)- {\rm h.c.}\right] \\
\dot{\delta c_{L}} & = & \left(-i\Delta'_{L}-\frac{\gamma_{\rm at}}{2}\right)\delta c_{L}- i G_{\rm{eff}}\left(\delta b+ \delta b^{\dag}\right)+ \sqrt{\gamma_{\rm at}}c_{L}^{in}\nonumber\\
\label{cl}& &+i N\bar{g}^2\frac{J}{z}\delta c_{R} -\bar{g}\sqrt{N}\left(\frac{J}{z}a_{R}^{in}+ \frac{y}{z}a_{L}^{in}\right)\sqrt{\gamma_{\rm c}}\\ 
\dot{\delta c_{R}} & = & \left(-i\Delta'_{R}-\frac{\gamma_{\rm at}}{2} \right)\delta c_{R}+ i \bar{G}_{\rm{eff}}\left(\delta b+ \delta b^{\dag}\right)+ \sqrt{\gamma_{\rm at}}c_{R}^{in}\nonumber\\
\label{cr}& & +iN\bar{g}^2\frac{J}{z}\delta c_{L}-\bar{g}\sqrt{N}\left(\frac{J}{z}a_{L}^{in}+ \frac{x}{z}a_{R}^{in}\right)\sqrt{\gamma_{\rm c}}
\end{eqnarray}
where 
\begin{eqnarray}
\Lambda & = &  g_{1}^2\alpha^{2} \Re\left({\frac{x+y-2J}{z}}\right), 
\;\;\tilde{\omega}_{m}  =  \omega_{m}+\Lambda\;,\nonumber\\
G_{\rm{eff}} & = &  g_{1}\bar{g} \sqrt{N}\frac{\alpha}{z}\left(y-J \right)\;,\;\;\;
\bar{G}_{\rm{eff}}  =   g_{1}\bar{g} \sqrt{N}\frac{\alpha}{z}\left(x- J \right) \;,\nonumber\\
 \Delta'_{L}& =& \Delta_{L}-\bar{g}^2 N\frac{y}{z}, \;\;\Delta'_{R} = \Delta_{R}-\bar{g}^{2} N\frac{x}{z}\;,
\end{eqnarray} 
where we have chosen $\alpha_L=\alpha_R=\alpha$ and the decay rates of the two ensembles to be the same, i.e., $\gamma_{1}=\gamma_{2}=\gamma_{\rm at}$, while $\Re(.)$ denotes the real part of the relevant quantity. 

\section{Effective Hamiltonian}
To obtain an effective Hamiltonian, that would lead to the Langevin Eqs. (\ref{b})-(\ref{cr}), we next choose $\delta'€™_L=-\delta'€™_R$, such that the quantity $z$ can be replaced by its real part $z_R=-(\frac{1}{2}\gamma_c^2+\delta'^2_L+J^2)$. In the limit, $\delta_L,\delta_R\gg \gamma_c$, we obtain the following Hamiltonian:
\begin{eqnarray}
 H_{\rm{eff}} &  = &   \sum_{\alpha\in {\rm L,R}}\left(\Delta'^{\rm R}_{\alpha} \delta c_{\alpha}^{\dag} \delta c_{\alpha}\right)+ \tilde{\omega}_{m}\delta b^{\dag} \delta b+\left[G_{\rm{eff}}^{\rm R}\left(\delta c_{L}+\delta c_{L}^{\dag}\right)\right.\nonumber\\
 & & \left.- \bar{G}_{\rm{eff}}^{\rm R}\left(\delta c_{R}+ \delta c_{R}^{\dag}\right)\right]\left(\delta b + \delta b^{\dag}\right)+\frac{\Lambda}{2}\left(\delta {b^\dag}^2+ \delta b^2\right)\nonumber\\
& & - C\left(\delta c_{L}\delta c_{R}^{\dag}+\delta c_{R} \delta c_{L}^{\dag}\right)\;,
\label{finH}
\end{eqnarray}
 where
 \begin{eqnarray}
\Delta'^{\rm R}_{\rm L}& =& \Delta_{\rm L}-\bar{g}^2N\frac{\delta'_R}{|z_R|}\;\;,\nonumber\\
\Delta'^{\rm R}_{\rm R}& =& \Delta_{\rm R}-\bar{g}^2N\frac{\delta'_L}{|z_R|}\;,
\end{eqnarray} 
arising from the real parts of the complex detunings $\Delta'_{L,R}$, represent the Stark shifts of unperturbed energy of the respective ensembles, 
\begin{equation}
C=-\bar{g}^2N\frac{J}{|z_R|}
\end{equation}
is the strength of the direct coupling, mediated by the tunnelling, 
and the coefficient $\Lambda$ of the non-linear term gets simplified to $2\alpha^2g_1^2\frac{J}{|z_R|}$.
Further the effective coupling constants $G_{\rm eff}$ and $\bar{G}_{\rm eff}$ are replaced by their real parts as 
\begin{equation}
G^R_{\rm{eff}} =  g_{1}\bar{g} \sqrt{N}\frac{\alpha}{|z_R|}\left(\delta'_R+J \right),
\bar{G}^R_{\rm{eff}}  =  g_{1}\bar{g} \sqrt{N}\frac{\alpha}{|z_R|}\left(\delta'_L+J \right). 
\end{equation}

Note that for a choice of $\delta'_L=-\delta'_R$, the Hamiltonian becomes hermitian, if we choose $\delta_L,\delta_R\gg \gamma_c$. Further, this amounts to a modified dissipation rate of the atomic ensembles as 
\begin{equation}
\gamma_{\rm at}\rightarrow \gamma'_{\rm at}=\gamma_{\rm at}-\bar{g}^2N\frac{\gamma_c}{|z_R|}\;.
\end{equation}
Clearly, the effective rate of atomic decay is decreased, due to the coupling with the membrane and the cavity mode.

\subsection{Validity of adiabatic approximation}
Before proceeding further, we consider a suitable parameter regime, in which the above approximation becomes valid. Usually, to eliminate the cavity modes adiabatically, the decay rates of the cavity modes should be much larger than the other decay rates involved, i.e., $\gamma_c\gg \gamma_{\rm at}, \gamma_{m}$, and the corresponding detunings should also obey $|\delta'_{L,R}|\gg \Delta_{L,R}$ \cite{lugiato}. Under this condition, the cavity fluctuations $\delta a_L$ and $\delta a_R$ become negligible at steady state, while we expect that the coupling constant $C$ between the ensembles will lead to an oscillatory dynamics of the fluctuations $\delta c_L$ and $\delta c_R$. To verify this, we consider the Eqs. (\ref{beforead1})-(\ref{beforead2}), that are obtained {\it before} the adiabatic elimination of the cavity mode is performed. Considering the white noise terms, such that $\langle a_L^{\rm in}\rangle$, $\langle a_R^{\rm in}\rangle$, $\langle b^{\rm in}\rangle$, $\langle c_L^{\rm in}\rangle$, $\langle c_R^{\rm in}\rangle$ $=0$, we solve for the dynamics of the expectation values of the fluctuations (see Fig. 2). This clearly shows that, for a realistic set of parameters, the fluctuations in the cavity modes remain negligible in magnitude, thereby justifying the adiabatic approximation.  On the other hand, the oscillations in the fluctuations of the atomic ensembles signify a Rabi coupling between them. 

\begin{figure}[h]
\begin{center}
\includegraphics[trim=0 190 0 200,clip,width=9cm]{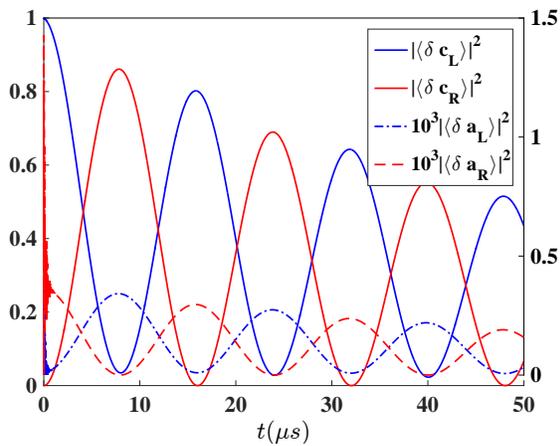}
\end{center}
\centering
\caption{Temporal variation of the expectation values $|\langle \delta c_L\rangle|^2$, $|\langle \delta c_R\rangle|^2$ (left y-axis) and $|\langle \delta a_L\rangle|^2$,$|\langle \delta a_R\rangle|^2$ (right y-axis) with time, normalized with respect to $2\pi\times 1$ MHz. The parameters chosen are $\gamma_L=\gamma_R=\gamma_c=2\pi\times 10$ MHz, $\gamma_1=\gamma_2=\gamma_{\rm at} = 2\pi\times 10$ kHz, $\gamma_m=2\pi\times1$ kHz, $\delta'_L=-\delta'_R=-2\pi\times 100$ MHz, $\omega_m=2J=2\pi\times 1$ GHz, $g_1=2\pi\times 1$ MHz, $\Delta_L=\Delta_R=2\pi\times 10$ MHz,  $\alpha_L=\alpha_R=\alpha=1$, and $\bar{g}\sqrt{N}=2\pi\times 10$ MHz. The initial condition is chosen as $\langle \delta c_L\rangle=1$, while all the other expectation values are zero.}
\end{figure}

Next we discuss the role of the tunnelling rate $J$ towards achieving a strong coupling between the ensembles. Specifically we consider two cases: a transparent membrane and a reflecting membrane. 

\subsection*{Case I: Transparent membrane ($J \neq 0$)}
If the membrane is transparent, the cavity modes $a_L$ and $a_R$ interact through their common coupling with the membrane mode. This can create a cross-talk between the two ensembles. In addition, as clear from the Hamiltonian (\ref{finH}), a possibility of direct coupling between the ensembles arises, with a coupling constant $C$. In the present case, suitable choices of parameters can make $C\gg |G_{\rm eff}|, |\bar{G}_{\rm eff}|$.

To demonstrate a large coupling in this case, we consider, as an example, the atomic ensembles on both the sides of the membrane, which interact with the respective cavity modes with an atom-field coupling  $\bar{g}\sqrt{N} =6.3$ MHz, while the cavity decay rate is $\gamma_c=2\pi \times 10$ MHz \cite{marquardt}. We emphasise that the two eigenmodes of the cavity interact resonantly via the common membrane mode, only if the difference in their frequencies follows the Raman resonance criterion $|\omega_+-\omega_-|\approx \omega_m$ \cite{law1}, where $\omega_{\pm}$ correspond to the optical normal modes $a_{\pm}=\frac{a_{L}\pm a_{R}}{\sqrt{2}}$, respectively. This leads to a non-adiabatic coupling between the two eigenmodes. In the present case, $|\omega_+-\omega_-|=2J$ and for a membrane with angular frequency $\omega_m=2\pi\times 1$ GHz, we obtain a required tunnelling rate $J=2\pi\times 500$ MHz.  Choosing $g_1=2\pi\times 1$ MHz, $\beta\approx 25$ (such that  $\delta_L=-\delta_R=-2g_1\beta\approx -2\pi\times 50$ MHz, satisfying the condition $\delta'_L=-\delta'_R$ as well) \cite{adia}, and $\alpha\rightarrow 0$, we find that $|C|\approx 2\pi\times 0.1$ MHz becomes much larger than $|G_{\rm eff}|$,  $|\bar{G}_{\rm eff}|$, and $|\Lambda|$.

Clearly, for realistic set-ups, the effective direct coupling between the two atomic ensembles becomes dominant, than any other coupling terms in the Hamiltonian. This leads to a non-adiabatic coupling between the ensembles with a time-scale of Rabi oscillation $1/C\sim 10^{-5}$ s (see Fig. 2), which is much smaller than the atomic decay time-scale $1/\gamma_{\rm at} \sim 10^{-4}$ s (specifically, if one chooses a metastable state as the excited state of the two-level atoms) - a clear indication of a strong coupling $C> \gamma_{\rm at}>\gamma'_{\rm at}$.  

\subsection*{Case II: Reflective membrane ($J \approx 0$)}
We next consider a case when the membrane is reflective i.e. $J\approx 0$. In this case, the modes $a_L$ and $a_R$ approximately represent the eigenmodes of the cavity, as we do not consider Raman resonance condition $2J=\omega_m$ any more \cite{bhatta-1}. 
We next choose the parameters such that a large Rabi coupling between the two atomic ensembles can be achieved via the mechanical mode bus, while the direct coupling strength $C$ becomes negligible. Accordingly, comparing the expressions of $G_{\rm eff}$, $\bar{G}_{\rm eff}$, and $C$ and using $|\delta'_L|\gg J$, we obtain the following condition: 
\begin{equation}g_1\alpha\gg \bar{g}\sqrt{N}\frac{J}{|\delta'_L|}\;.
\end{equation}
Choosing $g_1=2\pi\times 1$ MHz, $\bar{g}\sqrt{N}=2\pi\times 6.3$ MHz, $\delta_L=-2g_1\beta=-2\pi\times 50$ MHz, and $\beta=25$, the above condition reduces to $\alpha\gg 6.3\times 10^{-2}J/(2\pi)$, where $J$ is represented in MHz unit. Clearly, even for $J=2\pi\times 0.1$ MHz, $\alpha\sim 10$ should suffice to obtain a strong coupling between the ensembles. For a steady state value of $\alpha = 10$, the effective coupling constant then becomes $|G_{\rm eff}|\approx |\bar{G}_{\rm eff}|=\bar{g}\sqrt{N}g_1\alpha/|\delta'_L|=2\pi\times 0.63$ MHz. This leads to a Rabi oscillation between the two ensembles with a time-scale $|G_{\rm eff}|^{-1}\sim 1.58$ $\mu$s, which is much smaller than the time-scale of the atomic decay ($\sim 10^{-4}$ s) and that of the cavity decay ($\sim 10^{-7}$ s). 
\subsection{Stability of the cavity modes}
In the above analysis, we have adiabatically eliminated the cavity modes $a_L$ and $a_R$, to obtain an effective Hamiltonian. This means that in the steady state, the respective fluctuations $\delta a_L$ and $\delta a_R$ do not change substantially. However, the Hamiltonian (\ref{eq:H}), being intrinsically nonlinear, may lead to ever-increasing and large fluctuations at long times and thereby, to an instability in the cavity modes through their coupling to the membrane. This clearly violates the condition of adiabatic elimination of the cavity modes. Therefore, it is quite important to investigate whether, the system exhibits any stability at all, at least for certain relevant parameter domain, so that our analysis remains valid. 

We start with the Heisenberg equations of motion for the annihilation operators of the cavity modes, membrane, and the ensembles, obtained by using the Hamiltonian (\ref{eq:H}):
\begin{eqnarray}
\dot{a}_L&=&-i\left[\left(\delta_L-i\frac{\gamma_c}{2}\right)a_L-Ja_R-g_1a_L(b+b^\dag)\right.\nonumber\\
\label{al}&&\left.+\sqrt{N}\bar{g}c_L+\epsilon\right]\;,\\
\dot{a}_R&=&-i\left[\left(\delta_R-i\frac{\gamma_c}{2}\right)a_R-Ja_L+g_1a_R(b+b^\dag)\right.\;,\nonumber\\
\label{ar}&&\left.+\sqrt{N}\bar{g}c_R\right]\\
\label{bst}\dot{b}&=&-i\left[\left(\omega_m-i\frac{\gamma_m}{2}\right)b-g_1(a_L^\dag a_L-a_R^\dag a_R)\right]\;,\\
\label{calpha}\dot{c}_\alpha&=&-i\left[\left(\Delta_\alpha-i\frac{\gamma_{\rm at}}{2}\right) c_\alpha +\sqrt{N}\bar{g}a_\alpha\right]\;,\;\;\alpha\in L, R\;,
\end{eqnarray}
where the corresponding decay terms, $\gamma_c$, $\gamma_m$ and $\gamma_{\rm at}$, are phenomenologically included.
In the following, we focus on the fluctuation dynamics of the cavity modes only. For the membrane and the ensembles, we adopt a mean-field approach \cite{dong2014} so that the respective fluctuations can be negligible and the corresponding annihilation operators are replaced by their expectation values. Accordingly, we obtain the following in the steady state, using Eqs. (\ref{bst}) and (\ref{calpha}):
\begin{eqnarray}
\langle b\rangle &=&\frac{g_1}{\omega_m-i\gamma_m/2}\left(\langle a_L^\dag a_L\rangle - \langle a_R^\dag a_R\rangle\right)\;,\nonumber\\
\langle c_\alpha\rangle &=& -\frac{\sqrt{N}\bar{g}}{\Delta_\alpha-i\gamma_{\rm at}/2}\langle a_\alpha\rangle\;, \;\;\alpha\in L,R\;.
\end{eqnarray}
Replacing $a_\alpha \rightarrow \langle a_\alpha\rangle +\delta a_\alpha$ and the above equations in (\ref{al}) and (\ref{ar}), we finally have the following linearized coupled equations for the fluctuations of the cavity modes, written in a matrix form:
\begin{eqnarray}
\dot{\mathbf{F}}&=&\mathbf{MF}\;,\;\;\;\mathbf{F}=\left(\begin{array}{cc}\delta a_L & \delta a_R\end{array}\right)^T\;,\\
\mathbf{M}&=&\left(\begin{array}{cc}-i(\delta_L-r)-\frac{\gamma_c}{2} & iJ\\iJ & -i(\delta_R+r)-\frac{\gamma_c}{2}\end{array}\right)\;,\\
r&=&\frac{2g_1^2\omega_m}{\omega_m^2+(\gamma_m/2)^2}\left(\langle a_L^\dag a_L\rangle - \langle a_R^\dag a_R\rangle\right)\;,\nonumber
\end{eqnarray}
where $T$ represents the transpose of a matrix. If the matrix $\mathbf{M}$ has the eigenvalues with negative real parts, the fluctuations in the two cavity modes will decay to zero, leading to a stability in the dynamics at the steady state; otherwise, the fluctuations will exponentially increase with time, creating instability \cite{lugiato,dong2014,wu2011}.

In the present case, for $\delta_L=-\delta_R$, the eigenvalues of $\mathbf{M}$ can be simplified to 
\begin{equation}
\lambda_\pm = -\frac{\gamma_c}{2}\pm i\sqrt{J^2+(\delta_L-r)^2}\;,
\end{equation}  
displaying the negative real parts, irrespective of $J$, $\delta_L$, and $r$. Therefore, we conclude that the system under consideration is intrinsically stable against cavity fluctuation, in the chosen parameter regime, validating our analysis.

Note that this conclusion can be further verified, following \cite{lugiato}, by analyzing the eigenvalues of the linearized Eqs. (\ref{beforead1})-({\ref{beforead2}). We start by taking the expectation values of  these equations, as previously done in Sec. III A. The resultant equations can be written in a matrix form $\dot{\mathbf{A}}=\mathbf{NA}$, where $\mathbf{A}=[\langle \delta a_L\rangle, \langle \delta a_L^\dag\rangle,  \langle \delta a_R\rangle,  \langle \delta a_R^\dag\rangle,  \langle \delta b\rangle,  \langle \delta b^\dag\rangle,  \langle \delta c_L\rangle,  \langle \delta c_L^\dag\rangle,  \langle \delta c_R\rangle,$  $\langle \delta c_R^\dag\rangle ]$. We find that all the eigenvalues of the corresponding matrix $\mathbf{N}$ have negative real parts. Specifically, four eigenvalues can be considered as in a group, with the most negative real parts, while the corresponding eigenvectors span predominantly over $\langle \delta a_L\rangle$, $\langle \delta a_L^\dag\rangle$,  $\langle \delta a_R\rangle$,  $\langle \delta a_R^\dag\rangle$. The rest of the eigenvalues match with those obtained by similarly diagonalizing the Eqs. (\ref{b})-(\ref{cr}). This clearly obeys the conditions, required for stability against the fluctuations and adiabatic elimination of the cavity modes, as prescribed in Sec. VI of the Ref. \cite{lugiato}.

\section{conclusion}
In conclusions, we have presented a feasible scheme to obtain a strong coupling between two atomic ensembles, placed on both sides of an oscillating membrane suspended inside an optical cavity. The ensembles interact with the cavity mode on the respective side of the membrane. In the low excitation limit of the ensembles of $N (\gg 1)$ atoms and by adiabatically eliminating the cavity modes, we obtain an effective Hamiltonian for the interaction between the ensemble and the membrane. We have shown that the effective interaction strength between the two ensembles can be controlled by suitably choosing the tunnelling rate $J$ between the cavity modes. Specifically, in presence of a transparent membrane, a direct coupling between the ensemble can be established, while a reflective membrane would lead to a coupling that is mediated by the membrane. Moreover, the coupling strength can be further enhanced by increasing the number $N$ of atoms in the ensemble. The time-scale of oscillation remains much less than the time-scales of all the relevant decays. We provide the relevant numerical results to justify the adiabatic elimination. We show that for the relevant parameters, the fluctuations in the cavity modes do not diverge and therefore the system is dynamically stable. The present results pave the way towards quantum communication using mesoscopic systems, e.g., atomic ensembles and the oscillating membrane.

\end{document}